\documentstyle[prl,aps,amsfonts,twocolumn,times,tighten]{revtex}
\newcommand{\trace}{\mathop{\rm Tr}\nolimits}

\newcommand{\bra}[1]{\langle#1|}
\newcommand{\ket}[1]{|#1\rangle}

\newcommand{\identity}{\openone}
\newcommand{\be}{$$} 
\newcommand{\ee}{$$} 
\newcommand{\bea}{\begin{eqnarray*}}
\newcommand{\eea}{\end{eqnarray*}}
\begin{document}
\draft
\title{Optimizing Completely Positive Maps using Semidefinite Programming}
\author{Koenraad Audenaert\cite{KAmail} and
Bart De Moor\cite{BDMmail}}
\address{Katholieke Universiteit Leuven, 
Dept. of Electrical Engineering (ESAT-SISTA) \\
Kasteelpark Arenberg 10, 
B-3001 Leuven-Heverlee, Belgium
}
\maketitle
\begin{abstract}
Recently, a lot of attention has been devoted to finding physically realisable
operations that realise as closely as possible certain desired transformations 
between quantum states, e.g.\ quantum cloning, teleportation, quantum gates, etc.
Mathematically, this problem boils down to finding a completely positive trace-preserving
(CPTP) linear map that maximizes the (mean) fidelity between the map itself and the 
desired transformation. In this note we want to draw attention to the fact that this
problem belongs to the class of so-called semidefinite programming (SDP) problems. 
As SDP problems are convex, it immediately follows that they do not suffer from local optima.
Furthermore, this implies that the numerical optimization of the CPTP map can, and should, be 
done using methods from the well-established SDP field, as these methods exploit convexity
and are guaranteed to converge to the real solution. Finally, we show how the duality inherent to
convex and SDP problems can be exploited to prove analytically the optimality of a proposed
solution. We give an example of how to apply this proof method by proving the optimality
of Hardy and Song's proposed solution for the universal qubit $\theta$-shifter (quant-ph/0102100).
\end{abstract}
The basic problem considered by a number of authors \cite{app1,app2,fiur} is: 
what physically realisable quantum operation comes
closest to a given, but potentially unphysical, transformation between quantum states? 
The operation is most generally described by a linear map $\$$; the physical realisability
requires that the map is completely positive and trace-preserving (CPTP).
The desired transformation can be specified in a number of ways, for example by
enumerating all possible input-output pairs of pure states 
$\{\ket{\text{in},k},\ket{\text{out},k}\}$. The dimensions of the input and
output Hilbert spaces, ${\cal H}_{\text{in}}$ and ${\cal H}_{\text{out}}$, 
denoted $d_1$ and $d_2$, respectively, can in
general be different. The symbol $k$ labels the different pairs and can
either be discrete or continuous.

In the most commonly used formalism, the CPTP map $\$$ that is
to implement the transformation is represented by an 
operator $X$ acting on the Hilbert space 
${\cal H}_{\text{in}}\otimes{\cal H}_{\text{out}}$.
The requirements of complete positivity and trace preservation result
in the constraints
\bea
X&\ge&0 \\
\trace\mbox{}_{\text{out}} X &=& \identity_{\text{in}}.
\eea

The requirement that the map must implement the transformation as closely as
possible can be quantified by the mean fidelity $F$:
\be
F = \sum_k \bra{\text{out,k}} 
\$(\ket{\text{in},k}\bra{\text{in},k})\ket{\text{out,k}}.
\ee
The sum in this equation must be an integral with an appropriate measure for $k$
if $k$ is continuous. 
In terms of the operator $X$, the fidelity is given by
\be
F = \trace XR,
\ee
with
\be
R = \sum_k (\ket{\text{in},k}\bra{\text{in},k})^T
\otimes \ket{\text{out,k}}\bra{\text{out,k}}
\ee
The great virtue of this measure-of-goodness of the map
is that the fidelity is linear in the operator $X$.
In this way the problem has been formulated as an optimization
problem: 
\be
\text{(P):} \left\{
\begin{array}{l}
\text{maximize} \trace XR \\
X\ge 0 \\
\trace\mbox{}_{\text{out}} X = \identity_{\text{in}}
\end{array}
\right.
\ee
In general, optimization problem (P) cannot be solved analytically and
one must resort to numerical methods. Most authors try to solve (P) using ad-hoc iteration schemes
involving Lagrange multipliers.
Using these schemes, various useful results have been obtained.
However, in our view, the convergence properties of these schemes are
questionable, as it has not been proved that the solution obtained is actually
the global optimum. In fact, these methods reportedly get stuck now and then
in suboptimal local optima \cite{perscomm}.

In this note we wish to draw attention to the fact that problem (P) belongs
to a well-studied class of optimization problems called semidefinite programs (SDP). 
The importance of this fact cannot be overestimated. 
First of all, semidefinite programs
are a subclass of the class of convex optimization problems, and convex
problems have the very desirable property that a local optimum is
automatically a global optimum. Keeping this in mind we see that the
reported presence of local optima in the above iteration schemes is due
to the scheme itself, and not to the problem being solved.

Secondly, very efficient numerical methods have been devised to solve SDPs,
as these problems occur over and over again in various engineering disciplines, operations
research, etc.
These methods have very good convergence properties, and, moreover, they yield
numerical intervals within which the solution must lie. Using a sufficient number of
iterations, the width of this interval can be made arbitrarily small (apart from
numerical errors and given the validity of some technical requirements). 
In other words: convergence to the real solution is almost always guaranteed.
This is to be contrasted with ordinary methods, which typically yield one
outcome only, and it is difficult to know how far its value is removed from
the real solution, especially when the optimization problem has multiple local optima.

Thirdly, the way in which these numerical methods work can be exploited
to prove analytically that a given proposed solution, e.g.\ an analytical
Ansatz based on an educated guess and on the outcome of numerical experiments, is
actually the correct solution.

In the rest of this section we will first discuss the basic mathematical facts of
semidefinite programming and then apply them to the problem at hand.
For a short introduction to the subject, we refer to \cite{lieven}, and for an in-depth
treatment to \cite{nn}.
Note that \cite{rains} presents another application of SDP to quantum mechanics, namely
to finding bounds on the distillable entanglement of mixed bipartite quantum states.

The basic SDP problem is the minimization of a linear function of a real variable
$x\in R^m$, subject to a matrix inequality:
\be
\begin{array}{l}
\text{minimize } c^T x \\
F(x)=F_0+\sum_{i=1}^m x_i F_i\ge 0
\end{array}
\ee
where the $\ge$-sign means that $F(x)$ is positive semidefinite (hence the term SDP).
The problem data are the vector $c\in R^m$ and the $m+1$ real symmetric matrices $F_i$.
Alternatively, the $F_i$ can also be complex Hermitean but this is an atypical formulation 
within the SDP community (in engineering one typically deals with real quantities). 

This problem is called the {\em primal} problem. Vectors $x$ that satisfy the constraint
$F(x)\ge 0$ are called {\em primal feasible points}, and if they satisfy $F(x)>0$ they are called
{\em strictly feasible points}. The minimal objective value $c^T x$ is by convention denoted as $p^*$
(no complex conjugation!) and is called the {\em primal optimal value}.

Of paramount importance is the corresponding {\em dual} problem, associated to the primal one:
\be
\begin{array}{l}
\text{maximize } -\trace F_0 Z \\
Z\ge 0 \\
\trace F_i Z=c_i, \; i=1..m
\end{array}
\ee
Here the variable is the real symmetric (or Hermitean) matrix $Z$, and the data $c,F_i$ are the
same as in the primal problem. Correspondingly, matrices $Z$ satisfying the constraints are
called {\em dual feasible} (or {\em strictly dual feasible} if $Z>0$).
The maximal objective value $-\trace F_0 Z$, the {\em dual optimal value}, is denoted as $d^*$.

The objective value of a primal feasible point is an upper bound on $p^*$, 
and the objective value of a dual feasible point is a lower bound on $d^*$.
The main reason why one is interested in the dual problem is that one can prove that, under relatively 
mild assumptions, $p^*=d^*$. This holds, for example, if either the primal problem or the dual
problem are strictly feasible, i.e.\ there either exist strictly primal feasible points or
strictly dual feasible points. If this or other conditions are not fulfilled,
we still have that $d^*\le p^*$.
Furthermore, when both the primal and dual problem are strictly feasible, one proves
the following optimality condition on $x$:
$x$ is optimal if and only if $x$ is primal feasible and there is a dual feasible $Z$ such that
$ZF(x)=0$. This latter condition is called the {\em complementary slackness} condition.

In one way or another, numerical methods for solving SDP problems always exploit the inequality
$d\le d^*\le p^*\le p$, where $d$ and $p$ are the objective values for any dual feasible point
and primal feasible point, respectively. The difference $p-d$ is called the duality gap, and the 
optimal value $p^*$ is always ``bracketed'' inside the interval $[d,p]$. These numerical methods
try to minimize the duality gap by subsequently choosing better feasible points. Under the requirements
of the above-mentioned theorem, the duality gap can be made arbitrarily small (as far as numerical
precision allows). This is precisely the reason why one should be happy when an optimization
problem turns out to be an SDP problem.

We now apply these generalities to our problem at hand. Problem (P) can immediately be rewritten
as a (primal) SDP problem by noting that the set of Hermitean matrices form a {\em real} vector space
of dimension the square of the matrix dimension. Since we are dealing with matrices over the bipartite
Hilbert space ${\cal H}_{\text{in}}\otimes{\cal H}_{\text{out}}$ it is convenient to choose the basis
vectors of the matrix space accordingly. Let $\{\sigma^j\}$ and $\{\tau^k\}$ be orthogonal bases for Hermitean matrices over ${\cal H}_{\text{in}}$ and ${\cal H}_{\text{out}}$, respectively, then
$\{\sigma^j\otimes\tau^k\}$ forms an orthogonal basis for 
${\cal H}_{\text{in}}\otimes{\cal H}_{\text{out}}$.
Furthermore, choose the bases so that both $\sigma^0$ and $\tau^0$ are the identity matrix (of appropriate
dimension) and all other $\sigma^j$ and $\tau^k$ are traceless Hermitean matrices. An obvious choice
would be the set of Pauli matrices $\{\sigma^x,\sigma^y,\sigma^z\}$ or generalisations thereof to 
higher dimensions.
We thus have the following parameterisation of the matrix $X$:
\be
X = \sum_{j=0}^{d_1^2-1} \sum_{k=0}^{d_2^2-1} x_{jk}\sigma^j\otimes \tau^k.
\ee
With this parameterisation, the TP requirement can be expressed in a straightforward way. The
condition $\trace\mbox{}_{\text{out}} X = \identity_{\text{in}} = \sigma^0$ is fulfilled if and
only if $x_{j0}=0$ for all $j>0$, and $x_{00}=1/d_2$. By changing the parameterisation of $X$,
this can be taken care of implicitly:
\be
\begin{array}{rcl}
X &=& \sum_{j=1}^{d_1^2-1} \sum_{k=1}^{d_2^2-1} x_{jk}\sigma^j\otimes \tau^k \\ 
&& + \sum_{k=1}^{d_2^2-1} x_{0k}\sigma^0\otimes \tau^k \\
&& + \identity/d_2.
\end{array}
\ee
From this parameterisation, and the additional requirement $X\ge0$, it immediately follows
that the matrices $F_i$ (in the SDP problem) are given by
\bea
F_0 &=& \identity/d_2 \\
F_{\text{``$i$''}} &=& \sigma^j\otimes \tau^k,\text{ with }k\neq 0.
\eea
The index ``$i$'' in the left-hand side refers to the
$i$ of the SDP problem, and corresponds to all possible pairs $(j,k)$ of right-hand side indices
with $k\neq 0$. As a shorthand for summation over all these pairs we will use the symbol
$\sum_{j,k}^*$.

Finally, we can assign values to the vector coefficients $c_i$ as follows.
The fidelity $F$ is to be maximized, so we need an additional minus sign; furthermore,
in terms of $x_{jk}$, $F$ equals 
\be
F=\sum_{j,k}^* x_{jk} \trace(\sigma^j\otimes\tau^k R) + 1/d_2,
\ee
where we have used the fact that $\trace R=1$.
This yields for the coefficients $c_i$:
\be
c_{\text{``$i$''}} = -\trace(\sigma^j\otimes\tau^k R),
\ee
and for the optimal fidelity, in terms of the primal optimal value:
\be
F_{\text{opt}} = -p^* +1/d_2.
\ee

Using these expressions for the vector $c$ and the matrices $F_i$ (which are only dependent
on the dimensions of the problem!), one can go about solving the problem (P) numerically.
As some of the $F_i$ are complex, one has to use SDP software that explicitly allows complex
entries (e.g.\ \cite{sturm}).

Using the above assignments, the dual problem can now be formulated in a rather nice way.
The dual objective, to be maximized over all $Z\ge0$, is
\be
d=-\trace F_0 Z = -\trace Z/d_2.
\ee
The constraint $\trace F_i Z=c_i$ gets an interesting form:
\be
\trace(\sigma^j\otimes \tau^k(Z+R))=0,\text{ with }k\neq0.
\ee
As $Z$ and $R$ are both Hermitean, this means that the matrix $Z+R$ must be of the form
$Z+R = a_0\identity+\sum_{j\neq0} a_j \sigma^j\otimes\identity$,
or, in other words, 
\be
Z=a_0\identity+A\otimes\identity-R,
\ee 
with $A$ a traceless Hermitean matrix.
With this parameterisation for all dual feasible $Z$, the dual objective
becomes
\be
d=-d_1 a_0+1/d_2.
\ee
Maximizing $d$ thus amounts to minimizing $a_0$ over all traceless Hermitean matrices $A$
such that the resulting $Z$ is still positive semidefinite. From the parameterisation
of $Z$ one sees that the smallest feasible value of $a_0$ for a fixed matrix $A$ is given by
\be
a_0(A) = -\lambda_{\text{min}}(A\otimes \identity-R),
\ee
where $\lambda_{\text{min}}$ signifies the minimal eigenvalue of the matrix.
The dual problem finally becomes:
find the optimal traceless Hermitean matrix $A$ such that this $a_0(A)$ is minimal.
The dual optimal value is then 
\be
d^* = -d_1 \min_A a_0(A)+1/d_2.
\ee
Note that we have significantly reduced the number of unknown parameters: from $(d_1d_2)^2$ for $Z$
to $d_1^2-1$ for $A$.

These expressions for the primal and dual problem can be used for proving that a
certain proposed solution is optimal. To that purpose one needs to propose primal and dual
feasible points $x$ and $A$; if the resulting primal and dual objective values $p$ and $d$
turn out to be equal to each other, then $x$ and $A$ are optimal feasible points and $p=d=p^*=d^*$.
Alternatively, any feasible choice for $x$ and $A$ gives upper and lower bounds on the optimal
value $p^*$, resulting in lower and upper bounds, respectively, for the fidelity of problem (P).
For example, setting $A=0$ gives $a_0(A) = \lambda_{\text{max}}(R)$ resulting in the upper bound
$F\le d_1\lambda_{\text{max}}(R)$, which was already derived in \cite{fiur}.

Using the method of the previous paragraph, one can test whether the feasible points are optimal
or not, but it does not solve the problem of finding these points. As there is no hope for solving
the primal and dual problems analytically for all but the simplest problems, one must resort
to numerical methods. Luckily, efficient methods abound and some implementations are freely
available on the web. From the numerical results one can then try to guess the analytical form
of the solution, or at least try to propose an Ansatz containing a few unknown parameters.
If the number of parameters is small they could be found by solving the primal and dual problem
using the Ansatz.

Even this could be relatively complicated, especially for the dual problem, as this is an
eigenvalue problem.
An alternative for solving the dual problem is offered by the complementary slackness (CS) condition,
which does not require solving an eigenvalue equation. Supposing that a correct guess
has been made for $X$ of the primal problem, one then has to solve the linear equation
\be
(a_0\identity+A\otimes\identity-R)X=0
\ee
in the unknowns $a_0$ and $A$.
Of course, one then still has to prove that the resulting $Z$ is dual feasible, i.e.\ is positive
semidefinite, and this could still require solving an eigenvalue problem.

As an example of this proof technique, we now consider the problem of constructing an optimal 
qubit $\theta$-shifter, first considered by Hardy and Song \cite{hardy} and prove that their
``quantum scheme'' shifter (see also \cite{fiur}) is optimal.

A qubit $\theta$-shifter is a device that transforms a pure state 
$\psi(\theta,\phi) = \cos(\theta/2)\ket{0}+\exp(i\phi)\sin(\theta/2)\ket{1}$
into another pure state $\psi(\theta+\alpha,\phi)$. This is a non-physical operation and has,
therefore, to be approximated. Hardy and Song consider both a universal approximated shifter,
with fidelity independent of $\theta$, and a shifter with $\theta$-dependent fidelity
optimizing the {\em mean} fidelity. The mean fidelity of the non-universal shifter is better than for
the universal one, but it has only been proven for values of $\alpha$ equal to integer multiples 
of $\pi/2$ that it has {\em optimal} mean fidelity \cite{fiur}. We will now prove
optimality for {\em all} values of $\alpha$.

The matrix $R$ for the shifter is given by
\be
R=\left[
\begin{array}{cccc}
r_1 & 0   & 0   & r_5 \\
0   & r_2 & 0   & 0   \\
0   & 0   & r_3 & 0   \\
r_5 & 0   & 0   & r_4
\end{array}
\right],
\ee
with
\be
\begin{array}{rcl}
r_1&=&1/4+c-s \\
r_2&=&1/4-c+s \\
r_3&=&1/4-c-s \\
r_4&=&1/4+c+s \\
r_5&=&2c
\end{array}
\mbox{ and }
\begin{array}{rcl}
c&=& \frac{1}{12}\cos\alpha \\
s&=& \frac{\pi}{16}\sin\alpha.
\end{array}
\ee


The Ansatz for the primal feasible point is \cite{fiur}
\be
X=\left[
\begin{array}{cccc}
\cos^2\beta & 0   & 0   & \cos\beta \\
0   & \sin^2\beta & 0   & 0   \\
0   & 0   & 0 & 0   \\
\cos\beta & 0   & 0   & 1
\end{array}
\right].
\ee
There appear to be two regimes, depending on the value of $\alpha$.
For $\alpha\le\alpha_0=\arctan(8/3\pi)$, put $\cos\beta=1$,
and for $\alpha\ge\alpha_0$, $\cos\beta=c/(s-c)$.
This gives as primal objective fidelity
\bea
F &=& (1+\cos\alpha)/2, \text{ for }\alpha\le\alpha_0 \\
F &=& 1/2+2s+2c^2/(s-c), \text{ for }\alpha\ge\alpha_0.
\eea

Going over to the dual problem, we now present our own Ansatz for the dual feasible point $A$,
which was inspired by numerical results: consider diagonal $A$ only.
This means that $A$ is parameterised by a single number, say $\delta$, and equals 
$A=\delta\sigma^z$. This gives for $Z$:
\be
Z=\left[
\begin{array}{cccc}
a_0+\delta-r_1 & 0   & 0   & -r_5 \\
0   & a_0+\delta-r_2 & 0   & 0   \\
0   & 0   & a_0-\delta-r_3 & 0   \\
-r_5 & 0   & 0   & a_0-\delta-r_4
\end{array}
\right].
\ee
To prove optimality of both Ansatzes, we use the complementary 
slackness condition (for finding the optimal value for $a_0$ and $\delta$).
The CS condition $ZX=0$ gives rise to just three independent equations:
\bea
(a_0+\delta-r_1)\cos\beta-r_5&=&0 \\
(a_0+\delta-r_2)\sin^2\beta&=&0 \\
(a_0-\delta-r_4)-r_5\cos\beta&=&0
\eea
As could be expected, there are two different solutions:
\bea
a_0&=&1/4+3c \\
\delta&=&-s \\
\cos\beta&=&1
\eea
and
\bea
a_0&=&1/4+s+c^2/(s-c) \\
\delta &=& -s/(s-c) \\
(r_2-r_1)\cos\beta&=&r_5.
\eea 
The third equation of each set shows us that the first solution
pertains to the case $\alpha\le \alpha_0$
and the second solution to the other case.
The first solution gives mean fidelity 
\be
F=2a_0=(1+\cos\alpha)/2,
\ee
and the second solution 
\be
F=1/2+2s+2c^2/(s-c).
\ee
These values are exactly the ones obtained in the primal problem, so this
proves the optimality of our Ansatzes, provided $Z\ge0$ in both cases.
It is a basic exercise in linear algebra to calculate the eigenvalues of $Z$
in both cases; noting that $0\le s\le c$ in the case $\alpha\le\alpha_0$,
and $c\le s$ in the other case, one can indeed show that $Z$ is always positive 
semidefinite, proving its feasibility.

To conclude, we have noted that the problem (P), which has to be solved for finding
CPTP maps that optimally approximate certain desired qubit-transformations,
is a semidefinite programming (SDP) problem. From this observation, it follows
that (P) can be efficiently solved using standard SDP software, and that
there is no need for ad-hoc solution methods, which could suffer from 
bad convergence properties. Furthermore, we presented a method for proving
analytically that an Ansatz for the solution of (P) is optimal.
We hope that the present work will be useful for those working in the field
of determining optimal CP maps or optimal quantum measurements.

This work has been supported by the IUAP-P4-02 program of the Belgian state.


\begin{thebibliography}{9}
\bibitem[*]{KAmail} koen.audenaert@esat.kuleuven.ac.be
\bibitem[\dagger]{BDMmail} bart.demoor@esat.kuleuven.ac.be
%
\bibitem{app1} M. Sacchi, Phys.Rev.A {\bf 63}, 054104 (2001).
\bibitem{app2} S. Massar and S. Popescu, Phys.Rev.A {\bf 61}, 062303 (2000).
\bibitem{perscomm} J. Fiurasek, personal communication (2001).
\bibitem{lieven} L. Vandenberghe and S. Boyd, Semidefinite Programming, SIAM Review 38, 49-95 (1996).
\bibitem{nn} Y. Nesterov and A. Nemirovsky, {\em Interior-point polynomial methods in convex programming},
vol. 13 of Studies in Applied Mathematics, SIAM, Philadelphia PA (1994).
\bibitem{rains} E. Rains, quant-ph/0008047, to appear in the
November issue of IEEE Trans. Inform. Th. (2001).
\bibitem{fiur} J. Fiurasek, quant-ph/0105124 (2001).
\bibitem{hardy} L. Hardy and S. Song, Phys.Rev. A {\bf 63}, 032304 (2001).
L. Hardy and S. Song, quant-ph/0102100 (2001).
\bibitem{sturm} J.F. Sturm, Optimization Methods and Software 11-12 (1999) 625-653. 
Software available at {\tt http://members.tripodnet.nl/SeDuMi}
\end{thebibliography}
\end{document}